\documentstyle[12pt]{article}

\newcommand{\EQ}{\begin{equation}}
\newcommand{\EN}{\end{equation}}
\newcommand{\bea}{\begin{eqnarray}}
\newcommand{\ena}{\end{eqnarray}}

\renewcommand{\a}{\alpha}
\renewcommand{\b}{\beta}

\newcommand{\pa}{\partial}

\newcommand{\z}{\zeta}

\renewcommand{\l}{\lambda}

\newcommand{\x}{\chi}

\renewcommand{\o}{\omega}

\renewcommand{\z}{\zeta}

\begin{document}
 \def\bq{\begin{quote}}
\def\eq{\end{quote}}
\topmargin -1.2cm
\oddsidemargin 5mm

\renewcommand{\Im}{{\rm Im}\,}
\newcommand{\NP}[1]{Nucl.\ Phys.\ {\bf #1}}
\newcommand{\PL}[1]{Phys.\ Lett.\ {\bf #1}}
\newcommand{\NC}[1]{Nuovo Cimento {\bf #1}}
\newcommand{\CMP}[1]{Comm.\ Math.\ Phys.\ {\bf #1}}
\newcommand{\PR}[1]{Phys.\ Rev.\ {\bf #1}}
\newcommand{\PRL}[1]{Phys.\ Rev.\ Lett.\ {\bf #1}}
\newcommand{\MPL}[1]{Mod.\ Phys.\ Lett.\ {\bf #1}}
\renewcommand{\thefootnote}{\fnsymbol{footnote}}

\newpage
\begin{titlepage}
\begin{flushright}
IFUM 441/FT\\
February 1993 \\
\end{flushright}
\vspace{1.5cm}
\begin{center}
{\bf{{\large SOLITONS IN TWO--DIMENSIONAL TOPOLOGICAL}\\
{\large FIELD THEORIES}}}\\
\vspace{1cm}
{ S. PENATI} \\
\vspace{2mm}
{\em Dipartimento di Fisica dell'Universit\`{a} di Milano and} \\
{\em INFN, Sezione di Milano, Via Celoria 16, I-20133 Milano, Italy}\\
\vspace{0.8cm}
{ M. PERNICI} \\
\vspace{2mm}
{\em INFN, Sezione di Milano, Via Celoria 16, I-20133 Milano, Italy}\\
\vspace{0.8cm}
{ D. ZANON} \\
\vspace{2mm}
{\em Dipartimento di Fisica dell'Universit\`{a} di Milano and} \\
{\em INFN, Sezione di Milano, Via Celoria 16, I-20133 Milano, Italy}\\
\vspace{2cm}
{{\bf{ABSTRACT}}}
\end{center}
\bq
We consider a class of $N=2$ supersymmetric non--unitary theories
in two--dimensional Minkowski spacetime which admit
classical solitonic solutions. We show how
these models can be twisted into a topological sector
whose energy--momentum tensor is a BRST commutator.
There is an infinite number of degrees of freedom
associated to the zero modes of the solitons.
As explicit realizations of such models we discuss the BRST quantization of
a system of free fields, while in the interacting case we study $N=2$
complexified twisted Toda theories.

\eq
\vfill
\end{titlepage}
\renewcommand{\thefootnote}{\arabic{footnote}}
\setcounter{footnote}{0}
\newpage

In the past few years progress has been made in the study of
non--critical strings. In particular
the $c<1$ theories are understood by now in terms
of rather different techniques, i.e. conformal field theory \cite{b1},
matrix models \cite{b2}, topological field theory \cite{b3}. In the
topological field theory framework one shows that correlation functions of the
observables are independent of the two--dimensional metric. Therefore
no modes propagate on the world--sheet manifold and the only relevant
quantities
are the zero--modes of the instantons. A distinguished class of these models
is described in euclidean space by $N=2$ supersymmetric lagrangians
which can be twisted into a topological sector using the $U(1)$ invariance of
the theory.
It has been shown that these systems coupled to topological gravity
correspond to string theories with $c\leq 1$ \cite{b3,b4,b7}.

In this letter we study a class of $N=2$ supersymmetric theories in
Minkowski spacetime and address the issue of their topological formulation. In
Minkowski space a theory can be twisted into a topological, BRST invariant
sector, if the Lorentz group can be combined with a $O(1,1)$ symmetry. We
show that in order to implement the $O(1,1)$ invariance one has to consider
models which have $N=2$ complexified supersymmetries, and this in turn
can be realized only in a non--unitary lagrangian description. However the
problem of non--unitarity is naturally solved in the topological twisted
version of the theory. In fact the existence of a BRST charge provides the
standard mechanism which allows to eliminate ghost--like fields from the
physical spectrum. The topological symmetry enforces no propagation
on the two--dimensional worldsheet and the remnant degrees of freedom are given
by the zero modes which determine the quantum numbers
of the physical observables.   A simple,
completely solvable example of such theories is provided by a system of
free fields. We discuss the physical spectrum and find that all
particles excitations are cohomologically trivial.
There is a single massless field propagating in the target spacetime.
If the free fields are compactified, the winding solitons are the
only physical states.

Then we switch on the interaction and introduce a potential with an
infinite number of critical points. The resulting topological theories have
infinite degrees of freedom given by the zero modes of the solitons
which interpolate between different extrema of the potential. These systems
can be explicitly realized by $N=2$ complexified affine Toda theories,
which indeed are described by non--unitary lagrangians of the general form that
allows the topological Minkowskian twist. It has been shown that
affine complex Toda theories admit solitonic solutions \cite{b9,b10,b11}
which, in spite
of the non--unitarity of the action, have real energy and momentum. We
consider the corresponding topological sector and show that there survives
an infinite number of solitonic configurations that can be studied at the
quantum level in a weak coupling approximation.
These are the issues that we elucidate in this letter.

\vskip 15pt
First we set our notations and conventions:
$N=1$ superspace is parametrized
by two light--cone Minkowski coordinates $z$,
$\bar{z}$
\bea
z \equiv x^+ = \frac{1}{\sqrt{2}} (x^0 + x^1) \qquad \qquad
\bar{z}\equiv x^-= \frac{1}{\sqrt{2}}(x^0 - x^1)  \nonumber \\
\partial \equiv \partial_z = \frac{1}{\sqrt{2}} (\partial_0 + \partial_1)
\qquad \qquad \bar{\partial} \equiv \partial_{\bar{z}} = \frac{1}{\sqrt{2}}
(\partial_0 - \partial_1) \qquad \qquad \Box = 2 \partial \bar{ \partial}
\ena
and by two spinor coordinates $\theta$, $\bar{\theta}$, with
$\theta^{\ast}=-\theta$, $\bar{\theta}^{\ast}=\bar{\theta}$. Complex
conjugation for the product of spinors is defined as
$(\x \psi)^{\ast} = \psi^{\ast} \x^{\ast}$. We introduce covariant
spinor derivatives
\EQ
D = \pa_{\theta} + i\theta \pa  \qquad \qquad \bar{D} = \pa_{\bar{\theta}}
- i\bar{\theta} \bar{\pa}
\EN
which satisfy $\{D,\bar{D}\}=0$, $D^2=i\pa$ and
$\bar{D}^2=-i\bar{\pa}$. Finally, a generic $N=1$ superfield is denoted by
\EQ
\Phi = \phi +\frac{1}{\sqrt{2}} \theta \psi + \frac{1}{\sqrt{2}}
\bar{\theta} \bar{\psi} + \theta \bar{\theta} F
\EN

\vskip 15pt
We start by considering a class of field theories
in two--dimensional Minkowski spacetime
described in terms of $2n$ complex $N=1$ superfields, with general action
\EQ
S= \frac{1}{\b^2} \int d^2zd^2\theta \left[ K_{ij} D \Phi^{(-)}_i
\bar{D} \Phi^{(+)}_j + V(\Phi^{(+)}) + V(\Phi^{(-)}) + {\rm h.c.} \right]
\label{m1}
\EN
where $d^2z=dzd\bar{z}$, $d^2\theta=\bar{D}D$, $\b$ is a coupling constant
and $K_{ij}$ is an invertible, symmetric, constant $n \times n$ real matrix.
We emphasize that no reality conditions are imposed on the superfields and
the theory is not unitary.

In addition to the explicit $N=1$ supersymmetry the action in eq.(\ref{m1})
possesses a second supersymmetry.
The supersymmetry transformations on the fields are respectively
\bea
\delta \Phi_i^{(\pm)} &=& \left( \varepsilon_L Q_L^{+} +
\varepsilon_R Q_R^{+}
\right) \Phi_i^{(\pm)} \nonumber \\
\delta \Phi_i^{(\pm)} &=&
\left( \z_L Q_L^{-} + \z_R Q_R^{-} \right) J \Phi_i^{(\pm)}
\label{8}
\ena
where the charges
\bea
Q_L^+ &=& i(\pa_{\theta} -i\theta \pa) \qquad\qquad~ Q_R^+ =
{}~i(\pa_{\bar{\theta}} +i\bar{\theta} \bar{\pa}) \nonumber \\
Q_L^- &=& \pa_{\theta} +i\theta \pa=D \qquad~~~~ Q_R^- = ~\pa_{\bar{\theta}}
-i\bar{\theta} \bar{\pa}=\bar{D}
\label{8.5}
\ena
satisfy the supersymmetry algebra $(Q_L^{\pm})^2=i\pa$,
$(Q_R^{\pm})^2=-i\bar{\pa}$ and $\{Q^+,Q^-\}=0$. The operator $J$ acts on the
fields as $J\Phi_i^{(\pm)} = \pm \Phi_i^{(\pm)}$.
Unlike in standard $N=2$ theories,
the parameters $\varepsilon$, $\z$ which appear in eq.(\ref{8})
are complex in general, so that the action in eq.(\ref{m1}) is invariant
under a complexified $N=2$ algebra. This implies that the usual $U(1)$
invariance is extended to a $U(1) \otimes O(1,1)$.
This will play a crucial role in the definition of the twisted
theory and in the construction of the BRST charge.

The complex theories under consideration can be formulated in
$N=2$ superspace embedding the $N=1$ superfields into chiral and antichiral
$N=2$ supermultiplets \cite{b12}.
One introduces $N=2$ spinor coordinates $\theta_+$,
$\theta_-$, $\bar{\theta}_+$, $\bar{\theta}_-$ and corresponding covariant
spinor derivatives satisfying
\EQ
\{ D_+,\bar{D}_+\} =i\pa~~~~~~~~~~~~~~~~\{ D_-,\bar{D}_-\} =-i\bar{\pa}
\label{N1}
\EN
Chiral and antichiral superfields $\Psi$ and $\bar{\Psi}= \Psi^{\ast}$ are
subject to the constraints
\EQ
\bar{D}_{\pm} \Psi=0~~~~~~~~~~~~~~~~D_{\pm}\bar{\Psi}=0
\label{N2}
\EN
Then one defines new coordinates
\bea
&~&\theta_1= \frac{1}{2} (\theta_+ +\bar{\theta}_+)~~~~~~~~
{}~~~~~~~~\bar{\theta}_1= \frac{1}{2} (\theta_- +\bar{\theta}_-) \nonumber\\
&~&\theta_2= \frac{i}{2} (\theta_+ -\bar{\theta}_+)~~~~~~~~
{}~~~~~~~~\bar{\theta}_2= \frac{i}{2} (\theta_- -\bar{\theta}_-)
\label{N3}
\ena
so that given a $N=1$ superfield $\Phi(\theta_1, \bar{\theta}_1)$, the
chirality condition in eq.(\ref{N2}) allows to reconstruct a $N=2$
superfield
\EQ
\Psi(\theta_+, \theta_-, \bar{\theta}_+, \bar{\theta}_-)=
e^{-2\theta_1 \theta_2 \pa +2 \bar{\theta}_1 \bar{\theta}_2 \bar{\pa}}
\Phi(\theta_1 - i\theta_2, \bar{\theta}_1 - i \bar{\theta}_2)
\label{N4}
\EN
Applying this procedure to the $N=1$ superfields $\Phi_i^{(\pm)}$, one
defines the following $N=2$ chiral and antichiral superfields
\bea
&~&\Phi_i^{(+)} ~\rightarrow \Psi_i^{(+)}~~~~~~~~~~~~~~~~~~
\Phi_i^{(+) \ast} \rightarrow \bar{\Psi}_i^{(+)} \nonumber\\
&~&\Phi_i^{(-) \ast} \rightarrow \Psi_i^{(-)} ~~~~~~~~~~~~~~~~~~
\Phi_i^{(-)} ~\rightarrow \bar{\Psi}_i^{(-)}
\label{N5}
\ena
Thus the action in eq.(\ref{m1}) can be reexpressed in $N=2$ superspace
\EQ
S = \frac{1}{\b^2} \int d^2z d^4\theta~~ K_{ij} \left[ \Psi_i^{(+)}
\bar{\Psi}_j^{(-)}
+\Psi_i^{(-)} \bar{\Psi}_j^{(+)} \right]
+\frac{1}{\b^2} \int d^2z d^2 \theta~~ W(\Psi) +
\frac{1}{\b^2} \int d^2z d^2\bar{\theta}~~W(\bar{\Psi})
\label{N6}
\EN
where the superpotential $W$ is given by
\EQ
W(\Psi)= V(\Psi^{(+)}) + V(\Psi^{(-)})
\label{N6.5}
\EN

We address now the issue of twisting the theory into a topological sector.
To this end it is convenient to rewrite the action in components
and eliminate the auxiliary fields through the field equations
$F_i^{(\pm)} = - K^{-1}_{ij}\frac{\pa V}{\pa \phi_j^{(\mp)}}$
\bea
S &=& \frac{1}{\b^2}
\int d^2z \left\{ K_{ij} \left[ \pa \phi_i^{(-)} \bar{\pa}\phi_j^{(+)}
+ \frac{i}{2} \psi_i^{(-)} \bar{\pa} \psi_j^{(+)} -\frac{i}{2}
\bar{\psi}_i^{(-)} \pa \bar{\psi}_j^{(+)} \right]
- \frac{\pa V}{\pa \phi_i^{(+)}} K_{ij}^{-1}
\frac{\pa { V}}{\pa \phi_j^{(-)}} \right. \nonumber \\
&~~&~~~~~~\qquad \left. + \frac12 \bar{\psi}_i^{(+)} \psi_j^{(+)} \frac{\pa^2
V}
{\pa \phi_i^{(+)} \pa \phi_j^{(+)}}  +
\frac12 \bar{\psi}_i^{(-)} \psi_j^{(-)} \frac{\pa^2 { V}}
{\pa \phi_i^{(-)} \pa \phi_j^{(-)}} ~+~{\rm h.c.} \right\}
\label{9}
\ena
Then it is clear that these models
possess a $\Lambda \equiv O(1,1)$ symmetry
\bea
\delta_{\Lambda} \phi_i^{(\pm)} &=& 0 \nonumber \\
\delta_{\Lambda} \psi_i^{(\pm)} &=& \pm \omega \psi_i^{(\pm)} \nonumber \\
\delta_{\Lambda} \bar{\psi}_i^{(\pm)} &=& \mp \omega \bar{\psi}_i^{(\pm)}
\label{12}
\ena
where $\o$ is a real parameter. Since this internal symmetry group is
isomorphic to the Lorentz group $L$
\bea
\delta_L \phi_i^{(\pm)} &=& 0 \nonumber \\
\delta_L \psi_i^{(\pm)} &=& \frac{\l}{2} \psi_i^{(\pm)} \nonumber \\
\delta_L \bar{\psi}_i^{(\pm)} &=& - \frac{\l}{2} \bar{\psi}_i^{(\pm)}
\label{13}
\ena
new transformations can be defined corresponding to a twisted
Lorentz group $L' \sim (\Lambda \otimes L)_{{\rm diag}}$.
With the choice $\o = -\frac{\l}{2}$ one obtains
\bea
\delta_{L'} \phi_i^{(\pm)} &=& 0 \nonumber \\
\delta_{L'} \psi_i^{(-)} &=& \l \psi_i^{(-)} \nonumber \\
\delta_{L'} \psi_i^{(+)} &=& 0 \nonumber \\
\delta_{L'} \bar{\psi}_i^{(-)} &=& - \l \bar{\psi}_i^{(-)}
\nonumber \\
\delta_{L'} \bar{\psi}_i^{(+)} &=& 0
\label{13.5}
\ena
Therefore, with respect to the Lorentz group $L'$ the new spin assignment
on the spinors is
\bea
\psi_i^{(+)} &:& s'=0 \qquad \qquad \bar{\psi}_i^{(+)} ~:~~s'=0
\nonumber \\
\psi_i^{(-)} &:& s'=1 \qquad \qquad \bar{\psi}_i^{(-)} ~:~~s'=-1
\label{14}
\ena
(The twist corresponding to the choice $\o = \frac{\l}{2}$
would simply interchange the spin assignment between
$\psi^{(+)}$, $\bar{\psi}^{(+)}$
and $\psi^{(-)}$, $\bar{\psi}^{(-)}$.)
It is now possible to construct a nihilpotent linear combination of the
supersymmetry charges with $s'=0$.
Indeed let us consider the charge
\EQ
Q= \frac{1}{\sqrt{2}} \left[ Q^+_L -iQ^+_R + i(Q^-_L -i Q^-_R)J \right]
\label{14.5}
\EN
Using eq.(\ref{8.5}) we obtain the
following transformations on the component fields
\bea
\delta \phi_i^{(+)} &=& \eta[ i \psi_i^{(+)} +\bar{\psi}_i^{(+)}] \nonumber \\
\delta \phi_i^{(-)} &=& 0  \nonumber \\
\delta \psi_i^{(+)} &=& - 2\eta K_{ij}^{-1} \frac {\pa V}{ \pa \phi^{(-)}_j}
 \nonumber \\
\delta \psi_i^{(-)} &=&- 2\eta \pa \phi_i^{(-)} \nonumber \\
\delta \bar{\psi}_i^{(+)} &=& 2i\eta K_{ij}^{-1}
\frac {\pa V}{ \pa \phi^{(-)}_j}
 \nonumber \\
\delta \bar{\psi}_i^{(-)} &=&  -2i\eta \bar{\pa} \phi_i^{(-)}
\label{15}
\ena
With respect to the twisted Lorentz group the parameter $\eta$
has spin zero. Moreover $Q^2 =0$, $Q^{\dag} = Q$ so that the charge in
eq.(\ref{14.5}) represents the BRST operator of the twisted theory.
It can be rewritten as
\EQ
Q = \frac{1}{\b^2}
\int dx  \left[ \frac{i}{\sqrt{2}} K_{ij} (i\psi^{(+)}_i \pa \phi_j^{(-)} +
\bar{\psi}^{(+)}_i \bar{\pa} \phi_j^{(-)}) - \frac{1}{\sqrt{2}}(\psi_i^{(-)}
+ i\bar{\psi}_i^{(-)}) \frac{\pa V}{\pa \phi_i^{(-)}} + {\rm h.c.} \right]
\label{19}
\EN
The holomorphic component of the stress--energy tensor
\EQ
T =
-K_{ij} \pa \phi_i^{(-)} \pa \phi_j^{(+)} - \frac{i}{2} K_{ij}
\psi_i^{(-)} \pa \psi_j^{(+)} + {\rm h.c.}
\EN
satisfies the conservation equation
\EQ\bar{\pa} T = \pa \Theta
\label{16}
\EN
where the trace is given by
\EQ
\Theta =
\frac{\pa V}{\pa \phi_i^{(+)}} K_{ij}^{-1} \frac{\pa V}
{\pa \phi_j^{(-)}} + \frac12 \psi_i^{(+)}\bar{\psi}_j^{(+)}
\frac{\pa^2 V}{\pa \phi_i^{(+)} \pa \phi_j^{(+)}} + {\rm h.c.}
\EN
It is easy to check that the corresponding central charge is zero and
$T$, $\Theta$ can be written as BRST anticommutators
\bea
T &=& \left\{ Q, \frac{1}{2} K_{ij} \psi_i^{(-)} \pa \phi_j^{(+)}
+ {\rm h.c.} \right\} \nonumber \\
\Theta &=& \left\{ Q, -\frac{1}{2} \psi_i^{(+)}
\frac{\pa V}{\pa \phi_i^{(+)}} + {\rm h.c.} \right\}
\label{25}
\ena
The BRST structure of our twisted theory shares some similarities
with the one considered in Ref.\cite{b5}.
In both cases the topological invariance is
a consequence of eq.(\ref{25}), while the lagrangian is not $Q$--exact.

We notice that the twist can be performed even if we impose
reality conditions
$\Phi_i^{(\pm)\ast} = \Phi_i^{(\pm)}$ on the superfields in eq.(\ref{m1}).
In fact with this choice
the theory has not a genuine $N=2$ supersymmetry
with a $U(1)$ generator, but it is still non--unitary and it
has an extended superalgebra containing a
$O(1,1)$. On the other hand the conditions $\Phi_i^{(+)\ast} = \Phi_i^{(-)}$
and $K_{ij}$ positive definite restore unitarity but the $O(1,1)$ invariance
is lost.

Now we study the physical content of these topological theories.
The physical states are given by cohomology classes
of the BRST operator. The BRST transformations in eq.(\ref{15}) imply
that $\frac{\pa V}{\pa \phi_i^{(-)}}$ is $Q$--exact so that its expectation
value on physical
states vanishes. This condition is satisfied at the critical points of the
potential $V(\phi^{(-)})$ and it ensures the nihilpotency of the BRST
operator at the quantum level.
As a consequence of the fact that the stress--energy
tensor is $Q$--exact it follows that the correlation
functions of the observables are independent of the space--time coordinates.
Moreover since the theory
contains an equal number of bosons and ghost--like fields
all particle excitations decouple. This guarantees the absence
of negative norm states which appear at the lagrangian level
due to the fact that the kinetic term is not positive definite.
The physical spectrum is characterized by
observables associated to BRST invariant operators. These are the
topological charge ${\cal T}= \int_{M} dx \frac{\pa \phi}{\pa x}^{(+)}$,
where $M$ is the space line,
and the operator ${\cal O}(P^{(+)})$ which maps one critical point
into another, being $P^{(+)}$ the canonical momentum conjugated to
$\phi^{(-)}$. States in different topological sectors are
orthogonal.

The physical spectrum can be explicitly determined in a weak--coupling
expansion around a critical point. Deforming the background solution
into a nearby one $\phi^{(\pm)}_i + \delta \phi^{(\pm)}_i$
we obtain the linearized bosonic field equations
\EQ
K_{ij} \pa \bar{\pa} \delta \phi_j^{(\pm)} +
\frac{\pa^2 V}{\pa \phi_i^{(\mp)} \pa \phi_j^{(\mp)}} K^{-1}_{jl}
\frac{\pa^2 V}{\pa \phi_l^{(\pm)} \pa \phi_m^{(\pm)}}\delta \phi_m^{(\pm)} = 0
\EN
These equations are satisfied also by the $\psi_i^{(+)}$,
$\bar{\psi}_i^{(+)}$ fermions
which are then deformations of the bosonic fields.
Since the combination $i\psi_i^{(+)} + \bar{\psi}_i^{(+)}$ is the BRST
transform of $\phi_i^{(+)}$ one can use the BRST invariance of the theory
to deform continuously $\phi^{(+)}$--backgrounds with the same
topological charge.
Therefore one can perform a weak--coupling expansion
around one representative within each sector.

In order to compute the relevant
physical quantities one has to select a specific form of the
potential.
The simplest case is the free theory described by the lagrangian in
eq.(\ref{9}) where we set $V=0$, $\b =1$ and compactify the fields
on $M=S^{1}$ with periodic boundary conditions.
In this case the topological charge is trivial and we will show
that the only physical quantity is the momentum $p^{(+)} = \frac{1}{\pi}
\int_{S^1} dx P^{(+)}$, where $P^{(+)} = \frac12 \frac{\pa
\phi}{\pa t}^{(+)}$.
It is sufficient to analyze the
case of two real bosonic fields $\phi^{(\pm)}$
and their fermionic partners with metric $K=1$.
The fields admit a mode expansion
\bea
\phi^{(\pm)}_L &=& q^{(\pm)}_L + p^{(\pm)}_L ~(t+x) + \sum_{n<0} \frac{1}
{\sqrt{2\pi} |n|} \left[
a_n^{(\pm)} e^{in(t+x)} + {a_n^{(\pm)}}^{\dag} e^{-in(t+x)} \right]
\nonumber \\
\phi^{(\pm)}_R &=& q^{(\pm)}_R + p^{(\pm)}_R ~(t-x) + \sum_{n>0}\frac{1}
{\sqrt{2\pi} |n|} \left[
a_n^{(\pm)} e^{in(x-t)} + {a_n^{(\pm)}}^{\dag} e^{-in(x-t)} \right]
\nonumber \\
\psi^{(\pm)} &=& \frac{2^{\frac{1}{4}}}{\sqrt{\pi}} b_0^{(\pm)} +
\frac{2^{\frac{1}{4}}}{\sqrt{\pi}} \sum_{n<0} \left[ b_n^{(\pm)}
e^{in(t+x)} + {b_n^{(\pm)}}^{\dag} e^{-in(t+x)} \right] \nonumber \\
\bar{\psi}^{(\pm)} &=& i\frac{2^{\frac{1}{4}}}{\sqrt{\pi}} \bar{b}_0^{(\pm)} +
i\frac{2^{\frac{1}{4}}}{\sqrt{\pi}} \sum_{n>0} \left[ \bar{b}_n^{(\pm)}
e^{in(x-t)} + {(\bar{b}_n^{(\pm)})}^{\dag} e^{-in(x-t)} \right]
\label{29}
\ena
where $q_L^{(\pm)} = q_R^{(\pm)} \equiv \frac{q^{(\pm)}}{2}$,
$p_L^{(\pm)} = p_R^{(\pm)} \equiv \frac{p^{(\pm)}}{2}$.
The canonical commutation relations are
\bea
[q^{(\pm)},p^{(\mp)}] &=& \frac{i}{\pi} \qquad \qquad
[a_m^{(\pm)},{a_n^{(\mp)}}^{\dag}] = |n| \delta_{m,n} \qquad \quad
m,n \in Z-\{0\} \nonumber \\
\{ b_m^{(\pm)}, {b_n^{(\mp)}}^{\dag} \} &=& \delta_{m,n} \quad \qquad m,n \leq
0
\nonumber \\
\{ \bar{b}_m^{(\pm)}, {(\bar{b}_n^{(\mp)})}^{\dag} \} &=& \delta_{m,n} \qquad
\quad m,n \geq 0
\ena
and ${b_0^{(\pm)}}^{\dag} = b_0^{(\pm)}$, ${(\bar{b}_0^{(\pm)})}^{\dag} =
\bar{b}_0^{(\pm)}$.
Setting the potential to zero in eq.(\ref{19}) we obtain the BRST charge
\bea
Q &=& \frac{1}{\sqrt{2}} \int_{0}^{2\pi} dx \left[ -\psi^{(+)} \pa \phi^{(-)}
+ i\bar{\psi}^{(+)} \bar{\pa} \phi^{(-)} \right] \nonumber \\
&=& -\frac{2^{\frac{1}{4}}}{\sqrt{\pi}} \left[ 2\pi b_0^{(+)} p_L^{(-)} +
2\pi \bar{b}_0^{(+)}p_R^{(-)} + i\sqrt{2\pi} \sum_{n<0} \left(
b_n^{(+)} {a_n^{(-)}}^{\dag} - {b_n^{(+)}}^{\dag} a_n^{(-)} \right)
\right. \nonumber \\
{}~~~~~&~& \left.
+ i\sqrt{2\pi} \sum_{n>0} \left( \bar{b}_n^{(+)}
{a_n^{(-)}}^{\dag} - {(\bar{b}_n^{(+)})}^{\dag} a_n^{(-)} \right) \right]
\ena
The Fock vacuum is defined to be annihilated by $a_n^{(\pm)}$,
$b_n^{(\pm)}$ and $\bar{b}_n^{(\pm)}$, $ n \neq 0$ oscillators,
and it carries momentum $p^{(\pm)}$.

Let us show first that all particle excitations are cohomologically
trivial. We start by considering the left--moving sector.
The oscillators
satisfy the following BRST transformations
\EQ
[Q_L, a_n^{(+)}] = -inb_n^{(+)} \qquad \quad \{ Q_L, b_n^{(-)} \} = ia_n^{(-)}
\qquad n<0
\EN
where we have defined
\EQ
Q_L = i\sum_{n<0} \left(
{b_n^{(+)}}^{\dag} a_n^{(-)}- b_n^{(+)} {a_n^{(-)}}^{\dag}   \right)
\EN
Denoting by $P_0$ the projector on zero--particle states, the projector on
$N$--particle states is
\EQ
P_N = \frac{1}{N} \sum_{n<0} \left[ \frac{1}{|n|}
{a_n^{(+)}}^{\dag} P_{N-1} a_n^{(-)}
+ \frac{1}{|n|} {a_n^{(-)}}^{\dag} P_{N-1} a_n^{(+)} + {b_n^{(+)}}^{\dag}
P_{N-1} b_n^{(-)} +{b_n^{(-)}}^{\dag} P_{N-1} b_n^{(+)} \right]
\EN
It is easy to show that $P_N = \{ Q_L,
R_N \}$, where
\EQ
R_N = \frac{i}{N} \sum_{n<0} \frac{1}{|n|}
\left[ {(b_n^{(-)})}^{\dag} P_{N-1} a_n^{(+)} -
{(a_n^{(+)})}^{\dag} P_{N-1} b_n^{(-)} \right]
\EN
{}From the condition $Q_L \left. |{\rm phys}\right \rangle =0$
it follows that $\left. |{\rm phys}\right \rangle$ is
cohomologically equivalent to $P_0 \left. |{\rm phys}\right \rangle$.
Extending this analysis to the right--moving sector one shows
that all particle excitations are unphysical.

Now we focus on the zero modes. We define
\EQ
Q_0 = p^{(-)} \left[ b_0^{(+)} +\bar{b}_0^{(+)} \right]
\EN
and expand any state on the $b_0$--modes
\EQ
\left. |\Omega\right \rangle = \left. | \Omega_1 \right \rangle
 + b_0^{(+)}\left. | \Omega_2 \right \rangle + \bar{b}_0^{(+)}
\left. | \Omega_3 \right \rangle +b_0^{(+)}\bar{b}_0^{(+)}
\left. | \Omega_4 \right \rangle
\label{30}
\EN
The physical state condition implies
\EQ
p^{(-)} \left. | \Omega_1 \right \rangle = 0 \qquad \qquad p^{(-)}
\left. | \Omega_2 \right \rangle
= p^{(-)} \left. | \Omega_3 \right \rangle
\label{30.5}
\EN
so that if $p^{(-)} \neq 0$ the states in eq.(\ref{30}) are
cohomologically trivial (for instance the
quantity $b_0^{(+)}\bar{b}_0^{(+)}  \left. | \Omega_4 \right \rangle$
can be written as $-Q_0
\frac{1}{p^{(-)}} b_0^{(+)} \left. | \Omega_4 \right \rangle$).
Therefore the physical states are characterized by the conditions
$p^{(-)}=0$ and $p^{(+)}$ arbitrary. Since the potential is
trivial, the operator which maps the set of critical points into itself is
the translation operator $p^{(+)}$. The zero--modes $q^{(\pm)}$ can be
interpreted as the coordinates of a two--dimensional target space with
metric $\eta_{+-}=1$. We stress that this theory is not topological in
target space since deformations of $\eta_{\mu \nu}$ in the action
are not $Q$--exact.
Therefore propagation on target space is possible and in fact
the physical states $|p^{(+)}\rangle$
describe a massless bosonic field right--moving
on this space. In this respect  they resemble $D=2$ string
theories where a massless mode propagates on a
two--dimensional space--time (cfr. the critical $N=2$ string \cite{b8} or
the bosonic $c=1$ string \cite{b6}).

The analysis can be extended to the toroidal compactification of
the free model. In this case the bosonic fields take values on a circle
\EQ
\phi^{(\pm)} \equiv \phi^{(\pm)} + 2\pi R n^{(\pm)}
\EN
where the winding numbers $n^{(\pm)}$ are related to the topological charges
${\cal T}^{(\pm)} = 2\pi R n^{(\pm)}$.
The mode expansion of the fields is given in eq.(\ref{29}) where now the left
and right components of the zero modes are independent and $p_L-p_R$ is the
winding number operator.

The decoupling of the particle excitations can be proven as in
the previous example. In the zero--mode sector we define
\EQ
Q_0 = b_0^{(+)}p^{(-)}_L +\bar{b}_0^{(+)}p_R^{(-)}
\EN
Thus the conditions in eq.(\ref{30.5}) become
\EQ
p_{L,R}^{(-)} \left. | \Omega_1 \right \rangle = 0 \qquad \qquad p_R^{(-)}
\left. | \Omega_2 \right \rangle = p_L^{(-)}
\left. | \Omega_3 \right \rangle
\EN
Again one can easily show that the states are cohomologically trivial
whenever $p_L^{(-)} \neq 0$ or $p_R^{(-)} \neq 0$.
Therefore the physical states are characterized by
$p_{L,R}^{(-)}=0$ and $p_{L,R}^{(+)}= \frac{m}{2\pi R} \pm \frac{n^{(+)}}{2}
R$.
They describe solitons with winding number $n^{(+)}$ and momentum
$\frac{m}{\pi R}$.

Now we turn to the interacting theories and look for a potential which admits
an infinite number of critical points with solitonic configurations
interpolating between them. What we need is a $N=2$ complexified supersymmetric
sine--Gordon theory or a generalization of it. In fact a whole class of
models is available: they are the supersymmetric Toda theories
based on the affine superalgebra $A^{(1)}(n,n)$ \cite{b13,b11,b14}.
This superalgebra
has rank $r=2n$ and it admits a purely fermionic set of $2n+2$ roots,
which can be represented in terms of $2n$--dimensional complex vectors.
The roots $\vec{\a}_j$ with $j=1,\dots,2n$
realize the Cartan matrix of the $A(n,n-1)$ superalgebra, i.e.
\EQ
\vec{\a}_j^2 =0 \qquad \quad \vec{\a_i} \cdot \vec{\a}_j
= 2(-1)^{i+1} \delta_{j,i+1} + 2(-1)^i
\delta_{j,i-1}
\label{3}
\EN
and $\vec{\a}_{2n+1}=-\sum_{j=1}^n \vec{\a}_{2j-1}$,
$\vec{\a}_0 = -\sum_{j=1}^n \vec{\a}_{2j}$.
The Toda equations of motion for the $2n$ complex $N=1$
superfields $\Phi^a$
\EQ
2\bar{D}D \Phi^a + \sum_{i=0}^{2n+1} \a_i^a e^{\vec{\a}_i \cdot \vec{\Phi}} =0
\label{4}
\EN
can be derived from a real, non--unitary lagrangian \cite{b15}
\EQ
S = \frac{1}{\b^2} \int d^2 z d^2 \theta \left[ D\vec{\Phi} \cdot \bar{D}
\vec{\Phi} + \sum_{i=0}^{2n+1} e^{\vec{\a}_i \cdot \vec{\Phi}}~+~{\rm h.c.}
\right]
\label{5}
\EN
We now perform the field redefinitions introduced in Ref.\cite{b11}
\bea
&~& \Phi_i^{(+)} =\vec{\a}_i \cdot \vec{\Phi}  \qquad \quad
 \Phi_i^{(-)}=\vec{\a}_{\sigma(i)} \cdot \vec{\Phi}  \qquad~~ i~{\rm even}
\nonumber \\
&~&  \Phi_i^{(-)}=\vec{\a}_i \cdot \vec{\Phi}  \qquad \quad
 \Phi_i^{(+)}=\vec{\a}_{\sigma(i)} \cdot \vec{\Phi}  \qquad~~ i~{\rm odd}
\label{6}
\ena
where $\sigma(i) \equiv 2n+1-i$, $i=1,\dots,n$, generates
an automorphism $\a_i \rightarrow
\a_{\sigma(i)}$ of the
Dynkin diagram of the $A(n,n-1)$ superalgebra. In terms of the
new complex superfields the action in eq.(\ref{5}) takes the general form
in eq.(\ref{m1})
where
\EQ
V(\Phi^{(\pm)})=
\sum_{i=1}^n e^{\Phi_i^{(\pm)}}
 +e^{-\sum_{i=1}^n \Phi_i^{(\pm)}}
\label{7}
\EN
and $K_{ij}$ is the $n \times n$ matrix
\EQ
K=
\left(\begin{array}{cccccc}
1 & 1 & 1 & 1 & \cdots & ~ \\
1 & 0 & 0 & 0 & \cdots & ~ \\
1 & 0 & 1 & 1 & \cdots & ~ \\
1 & 0 & 1 & 0 & \cdots & ~ \\
\cdots & \cdots & \cdots & \cdots & \cdots & ~ \\
\cdots & \cdots & \cdots & \cdots & \cdots & ~
\end{array}\right)
\quad
K^{-1}=
\left(\begin{array}{cccccc}
0 & 1 & 0 & \cdots & ~ & ~ \\
1 & 0 & -1 & 0 & \cdots & ~ \\
0 & -1 & 0 & 1 & 0 & \cdots \\
{}~ & ~ & ~ & \cdots & ~ & ~ \\
{}~ & ~ & ~ & ~ & 0 & (-1)^n \\
{}~ & ~ & ~ & ~ & (-1)^n & (-1)^{n+1}
\end{array}\right)
{}~~~~~~~~~~~
\EN

First we discuss the existence of solitonic configurations which can be
determined setting the fermion fields
to zero and
searching for nontrivial, finite energy solutions of the bosonic
equations of motion. As observed in Ref.\cite{b11} it is convenient
to redefine the bosonic variables
\bea
\phi_{2j-1}^{(-)} &=& i(\xi_j^{(1)}-\xi^{(2)}_j) \qquad \qquad \quad
\phi_{2j-1}^{(+)} = i(\xi_{n+1-j}^{(2)} - \xi_{n+2-j}^{(1)})
\quad~ j=1,\dots , \left[ \frac{n+1}{2} \right] \nonumber \\
{}~~&~&~~~~~~~~~ \nonumber \\
\phi_{2j}^{(-)} &=& i(\xi_{n+1-j}^{(1)} - \xi_{n+1-j}^{(2)}) \qquad
{}~~~\phi_{2j}^{(+)} = i(\xi_j^{(2)} - \xi_{j+1}^{(1)})
\qquad \quad~~ j=1, \dots , \left[ \frac{n}{2} \right] \nonumber \\
{}~~~~~~~~~~~~
\label{21}
\ena
with $\xi_0^{(1)} = \xi_{n+1}^{(1)} =
-\sum_{j=1}^n \xi_j^{(1)}$, $\xi_0^{(2)} = \xi_{n+1}^{(2)} =
-\sum_{j=1}^n \xi_j^{(2)}$. It is easy to check that the bosonic lagrangian
reexpressed in terms of the $\xi$-fields becomes
\EQ
{\cal L}_{{\rm B}} = {\cal L}(\xi^{(2)}) - {\cal L}(\xi^{(1)})~+~{\rm h.c.}
\EN
where ${\cal L}$ is the Toda lagrangian for the bosonic $a_n^{(1)}$ theory
with imaginary coupling
\EQ
{\cal L}= \frac{1}{\b^2} \sum_{j=0}^n \left[ \frac12  \pa \xi_j \bar{\pa}
\xi_j +  e^{i(\xi_j -\xi_{j+1})} \right]
\label{an}
\EN
We can then borrow all the results that have been obtained on solitonic
solutions for the $a^{(1)}_n$ complex Toda theory \cite{b9,b10}.
It has been shown that
even though the fields are complex the solitons have real energy and momentum.
Since they describe finite energy configurations they satisfy the boundary
conditions $ \vec{\xi}(x=\pm \infty) \in 2\pi
\Lambda_{\omega}$ where $\Lambda_{\omega}$
is the weight lattice of the algebra. In particular the one--soliton static
solutions are
\EQ
\xi_j^{(a)} = i~ \log{\frac{(1+e^{\sigma_a x-\lambda}\o^j_a)}
{(1+e^{\sigma_a x-\lambda}\o^{j-1}_a)}} \qquad ~~~ a=1,\dots,n
\EN
where $\o_a \equiv e^{\frac{2\pi a}{n+1}i}$, $\sigma_a = 2\sqrt{2}
\sin{\frac{\pi a}{n+1}}$ and $\l$ is a complex parameter.
The solitons of the supersymmetric $A^{(1)}(n,n)$ models are
given by the configurations $(\xi^{(1)},
\xi^{(2)})$, with $\xi^{(1)}$ and $\xi^{(2)}$
solitonic solutions of the corresponding bosonic theory.

Now we perform the twist and concentrate on the BRST invariant theory.
{}From the general discussion we know that $\phi^{(-)}_i$ must be an extremum
of the potential $V(\phi^{(-)})$.
In terms of the $\xi$ fields introduced in eq.(\ref{21}) this gives
\EQ
\vec{\xi}^{~(1)} = \vec{\xi}^{~(2)}~~~~~~{\rm mod}~ 2\pi \Lambda_{\omega}
\label{s1}
\EN
Therefore the topological $A^{(1)}(n,n)$ theory has the same
solitonic spectrum of the $a^{(1)}_n$ theory.
The energy--momentum tensor of these configurations vanishes in agreement
with topological invariance.

In the weak coupling expansion around a solitonic background
the bosonic field deformations and the fermionic fields
$\psi^{(+)}_i$, $\bar{\psi}^{(+)}_i$ satisfy the linearized $a^{(1)}_n$ Toda
field equations. The normal modes of the solutions can be determined
explicitly \cite{b16,b17} since multi--solitonic configurations are
known \cite{b9,b10}.
Using the canonical BRST analysis \cite{b18} it is easy to show that the
particle excitations are cohomologically trivial, whereas the study of the
zero--mode sector requires a careful treatment with collective coordinates.
We intend to report on this in a future publication.

\end{document}